\documentstyle[12pt]{article}\topmargin=-0.1in\oddsidemargin=5mm

\begin{document}\def\p{\phi}\def\P{\Phi}\def\a{\alpha}\def\e{\varepsilon}
\def\be{\begin{equation}}\def\ee{\end{equation}}\def\l{\label}
\def\0{\setcounter{equation}{0}}\def\b{\beta}\def\S{\Sigma}\def\C{\cite}
\def\r{\ref}\def\ba{\begin{eqnarray}}\def\ea{\end{eqnarray}}
\def\n{\nonumber}\def\R{\rho}\def\X{\Xi}\def\x{\xi}\def\la{\lambda}
\def\d{\delta}\def\s{\sigma}\def\f{\frac}\def\D{\Delta}\def\pa{\partial}
\def\Th{\Theta}\def\o{\omega}\def\O{\Omega}\def\th{\theta}\def\ga{\gamma}
\def\Ga{\Gamma}\def\t{\times}\def\h{\hat}\def\rar{\rightarrow}
\def\vp{\varphi}\def\inf{\infty}\def\le{\left}\def\ri{\right}
\def\foot{\footnote}
\begin{flushright}
{\large hep-ph/0003074}
\end{flushright}
\begin{center}
{\Large Multiplicity distribution tails at high energies}\\
{\it E.A.Kuraev, J.Manjavidze\foot{Permanent address: Inst.of
Phys.,  Tbilisi, Georgia}, A.Sissakian\\ JINR, Dubna, Russia}
\end{center}

\begin{abstract}
The idea that the hard channels may dominate in the very high
multiplicity processes is investigated. Quantitative realization of
the `hard Pomeron', deep inelastic scattering and large-angle
annihilation mechanism combinations are considered in the pQCD frame
for this purpose.
\end{abstract}

\section{Introduction}\0

Investigation of the multiplicity distributions was popular since
seventies \C{kai}. The modern hadron theory based on the local QCD
Lagrangians \C{bfkl} and the experimental consequences was given in
the review papers \C{khose}. The very high multiplicity (VHM)
processes, as the attempt to get beyond this standard multiperipheral
hadron physics, was offered in \C{lesha}. It considered as a possible
physical program for LHC experiments.

It was shown in seventieth that  the multiperipheral kinematics
dominates inclusive cross sections.  Moreover, the
created particles spectra do not depend on $s$ at high energies in
the multiperipheral region:
$$
f(s,p_c)=2E_c\frac{d\sigma}{d^3
p_c}=\int\frac{d t_1 d t_2 s_1s_2\phi_1(t_1)\phi_2(t_2)} {(2\pi)^2
s(t_1-m^2)^2(t_2-m^2)^2},~s_1s_2=sE_{c\bot}^2, E_{c\bot}^2=m_c^2+
\vec{p}_{c\bot}^2.
$$
Here $s_1=(p_a+p_c)^2, s_2=(p_b+p_c)^2$, and $\phi_i(t_i)$ are the
impact factors of hadrons.  So the particle $c$ forgot the details of
its creation.  It was found experimentally that the ratio
\be
\frac{f(\pi^+p\to\pi_- +...)}{\sigma(\pi^+p)}
=\frac{f(K^+p\to\pi_- +...)}{\sigma(K_+p)}=
\frac{f(pp\to\pi_- +...)}{\sigma(pp)}
\l{1.1}\ee
is universal \C{risk}. This take place due to the two Pomeron
multiperipheral exchange providing the nonvanishing contribution in
the $s$ asymptotics to the cross section. It was implied that the
Pomeron intercept is exactly equal to one. Just this kinematics leads
to $c_m=\ga_m(c_1)^m$ (the correlators $c_m$ are introduced in
(\r{corr})), i.e.  to the KNO-scaling \C{kno}.

We begin with general analysis to formulate an aim of this paper. Let
$\s_n(s)$ be the cross section of $n$ particles (hadrons) creation at
the total CM energy $\sqrt{s}$.  We introduce the generating
function:
\be
T(s,z)=\sum_{n=1}^{n_{max}} z^n\s_n(s),~s=(p_1+p_2)^2>>m^2,~
n_{max}=\sqrt{s}/m.
\l{1}\ee
So, the total cross section and the averaged multiplicity will be:
\be
\sigma_{tot}=T(s,1)=\sum\sigma_n,~\sigma_{tot}\bar{n}=\sum n\sigma_n=
\le.\frac{d}{dz}T(s,z)\ri|_{z=1}.
\ee

At the same time
\be
\sigma_n=\f{1}{2\pi i}\oint\f{dz}{ z^{n+1}}T(s,z)=
\f{1}{2\pi i}\oint\frac{dz}{z}e^{(-n\ln z+\ln T(s,z))}.
\ee
The essential values of $z$ in this integral are defined by the
equation (of state):
\be
n=z\f{\pa}{\pa z}\ln T(z,s).
\l{3}\ee
Considering the tail, i.e. $n>>\bar n$, let us assume that one can
find such values of $n<<n_s$ at high energies $\sqrt s>>m$ that we
can neglect in (\r{1}) dependence on the upper boundary $n_{max}$.
This formal trick allows to consider $T(z,s)$ as the nontrivial
function of $z$.  Then the asymptotics over n ($n<<n_s$ is assumed)
is governed by the mostleft situated singularity $z_s$ of $T(z,s)$:
\be
\s_n(s)\propto e^{-n\ln z_c(n,s)},
\l{2}\ee
where $z_c(n,s)$ is the smallest solution of eq.(\r{3}). It is
important that
\be
z_s(n,s)\to z_c~{\rm at}~n\to\infty.
\l{4}\ee
One can put this method of asymptotic estimation in the basis of VHM
processes phenomenology.

We may distinguish following possibilities at $n\to\infty$:\\
1)$z_s=1$: $\s_n>O(e^{-n})$;\\
2)$z_s=\infty$: $\s_n<O(e^{-n})$;\\
3)$i<z_s<\infty$: $\s_n=O(e^{-n})$.

The asymptotics 1) assumes the condensation phenomena \C{kac}.
The asymptotics 2) belong to the multiperipheral processes
kinematics: created particles form jets moving in the CM frame with
different velocities along the incoming particles directions, i.e.
with restricted transverse momentum. The third type asymptotics is
predicted by stationary Markovian processes with the QCD jets
kinematics of the high transverse momentum particles creation. It is
evident that the $n$ asymptotics should be governed by largest among
1) - 3). Just under this idea we hope that we get beyond the
multiperipheral kinematics in the VHM region.

Additional information is coded in the expansion over binomial
moments $c_m(s)$:
\be
\ln T(s,z)=\sum\frac{(z-1)^m}{m!}c_m(s).
\l{corr}\ee
So, for instance, if all $c_m=0,~m>1$, then we have the Poisson
distribution: $$\sigma_n=\s_{tot}\frac{(\bar{n})^n}{n!}
e^{(-\bar{n})}.$$ But if $c_m=\ga_m(c_1)^m$, where $\ga_m$ is the
some constant, then the so called KNO scaling take place:
$$\bar{n}\s_n= \s_{tot}\Psi(n/\bar{n}).$$

The multiple production processes are typical inelastic reactions of
the initial kinetic energy dissipation into the particles mass.
Consequently, the mean multiplicity $\bar{n}(s)$ is the measure of
entropy ${\cal S}$ production at given energy. Experimentally
$\bar{n}(s)\leq \ln^2s<<n_{max}$. This testify to the incomplete
energy dissipation in the mostly probable channels of hadrons
production.

This phenomena is explained naturally by presence of the space-time
local non-Abelian symmetry constraints. So, it is known \C{zakh} that
there is not thermalization phenomena in the completely integrable
systems. But, at all evidence, the quantum Yang-Mills theory is not
completely integrable, i.e. it admits the dissipation, but
nevertheless the symmetry constraints play essential role.

It is natural to assume that ${\cal S}$ exceed its maximum if
$n>>\bar{n}(s)$. So, our essentially inelastic process is happened
so rapidly that the non-Abelian symmetry constraints becomes frozen.
On other hand, maximum of entropy ${\cal S}$ means that the final
state of the dissipation process is equilibrium.

Last one means relaxation of energy correlations, i.e. absence of the
macroscopical energy flows in the system, and the Gauss energy
spectrum of created particles. We would like to say that in
such a state one get to the VHM.

The aim of this article is to build the suitable mechanism of
maximal initial energy dissipation, where the correlations are
relaxed and the energy spectra are Gaussian.. The classification
1)-3) is mostly general and we will put it in the basis of
consideration. So, our main purpose is to show as the multiperipheral
kinematics transform in the VHM domain.

\section{Pomeron,DIS and Double-Logarithmic kinematics}\0

Let us consider process of type $2\to 2+n$ in different kinematics
$$
A(P_1)+B(P_2)\to A'(p_1')+B'(p_2')+h_1(k_1)+h_2(k_2)+\cdots +
h_n(k_n), s=(p_1+p_2)^2>>m^2.
$$
We will distinguish the peripheral,deep inelastic and large-angles
kinematical regions with different physical content.
First we build two light-like 4-momenta from the momenta of initial
particles $p_{1,2}=P_{1,2}-P_{2,1}m^2_{2,1}/s$ and present the
4-momenta of final particles in form:
\begin{equation}
p_1'=\alpha_1' p_2+\beta_1' p_1+p_{1\bot}'; a_\bot p_{1,2}=0;
p_2'=\alpha_2' p_2+\beta_2' p_1+p_{2\bot}';
k_i=\alpha_i p_2+\beta_i p_1+k_{i\bot}'.
\end{equation}
Sudakov's parameters $\alpha,\beta$ are not independent.The mass
shell conditions and the conservation low give the relations:
\begin{eqnarray}
s\alpha_1'\beta_1'=m_1^2+\vec{p}_1^{'2}=E_{1\bot}^2, \\ \nonumber
s\alpha_2'\beta_2'=E_{2\bot}^2,s\alpha_i\beta_i=E_{i\bot}^2,
\\ \nonumber
\vec{a}^2=-a_\bot^2>0;
\alpha_1'+\alpha_2'+\sum \alpha_i=1;
\\ \nonumber
\beta_1'+\beta_2'+\sum \beta_i=1.
\end{eqnarray}

Consider first the peripheral kinematics. For it is
characteristic the weak dependence of differential cross sections on
the center of mass (CM) total energy $2E=\sqrt{s}$,the strict
ordering of parameters $\alpha,\beta$
$$
1\approx \beta_1'>>\beta_1>>...>>\beta_n>>\beta_2'\sim\frac{m^2}{s};
\frac{m^2}{s}<<\alpha_1'<<\alpha_1<<...<<\alpha_n<<\alpha_2'\approx 1
$$
and the restrictiveness on the transvers momenta $|\vec{k}_i|\sim m$.
It corresponds to small emission angles moving along 3-momentum
$\vec{P}_1$
$$
\theta_i=\frac{|\vec{k}_i|}{E\beta_i}<<1,|\beta_i|>>|\alpha_i|,
$$
and the similar expression for particles moving in opposite direction
$|\beta_i|<<|\alpha_i|$. The central region
$|\alpha_i|\sim|\beta_i|\sim E_{i\bot}/E<<1$ corresponds to particles
of low energies moving at large angles(in cm frame).  The
differential cross section have the form:
\begin{eqnarray}
d\sigma_{2\to 2+n}&=&\frac{(2\alpha_s)^{2+n}}{16\pi^{2n}}
C_V^n\frac{d^2q_1}{q_1^2+m^2}\frac{d^2q_2}{(q_1-q_2)^2+\lambda^2}...
\\ \nonumber &\times&\frac{d^2q_{n+1}}{(q_n-q_{n+1})^2 + \lambda^2}
\frac{1}{q_{n+1}^2+\lambda^2}\frac{d\alpha_1}{\alpha_1}
\theta(\alpha_2-\alpha_1)... \\ \nonumber
&\times&\frac{d\alpha_n}{\alpha_n}
\prod_{i=1}^{n+1}(\frac{s_i}{m^2})^{2\alpha (q_i^2)} =
\frac{1}{q_{n+1}^2+\lambda^2}dZ_n,
\end{eqnarray}
where $C_V=3$ and we imply $q_i$ the two-dimensional euclidean
vectors,the 4-momentum of the $i$-th emitted particle (gluon)is
\begin{equation}
k_i=(\alpha_i-\alpha_{i+1})p_2+(\beta_i-\beta_{i+1})p_1+
(q_i-q_{i+1})_\bot=
-\alpha_{i+1}p_2+\beta_ip_1+(q_i-q_{i+1})_\bot;
\end{equation}
$s_i$ are the partial squares of invariant mass of nearest emitted
particles:
\begin{equation}
s_1=(p_1'+k_1)^2=s|\alpha_2|,s_{n+1}=(k_n+p_2')^2=
\frac{E_{n\bot}^2}{\alpha_n},
s_i=E_{\bot,i-1}^2\frac{\alpha_{i+1}}{\alpha_{i-1}},
s_1s_2...s_n=s E_{1\bot}^2...E_{n\bot}^2,
\end{equation}
and the trajectory of reggeized gluon is
$$
\alpha(q^2)=\frac{q^2\alpha_s}{2\pi^2}\int\frac{d^2k}
{(k^2+\lambda^2)((q-k)^2+\lambda^2}.
$$
Here $\lambda$ is gluon mass.It was shown that the infrared
singularities absent in the limit $\lambda\to 0$.In this point we
will suggest the gluon to be massive:
$\lambda\to M$ and will decay to the jet of hadrons (pions) with the
probability to create $n$ particles
$dW_n(M)=\frac{c}{\bar{n}}exp(-n/\bar{n})dn,\bar{n}=
\ln(M^2/m_{\pi}^2)$.

The Monte Carlo simulation shows a tendency to minimization of the
number of rungs at large $n$.

For the pure deep inelastic case, when one of the initial hadrons is
scattered at the angle $\theta$ have the energy $E'$ in the cms of
beams whereas the another is scattered at small angle and the large
transfer momentum $Q=4EE'\sin^2(\theta/2)>>m^2,$ is distributed to
the some number of the emitted particles due to evolution mechanism
we have \cite{basic}($\theta$ is small):
\begin{eqnarray}
d\s_n^{DIS}=\frac{4\alpha^2E^{{'}2}}{Q^4 M} dD_n dE'd\cos\th,
\n\\
dD_n=(\frac{\alpha_s}{4\pi})^n\int_{m^2}^{Q^2}
\frac{d k_n^2}{k_n^2}\int_{m^2}^{k_n^2}\frac{d
k_{n-1}^2}{k_{n-1}^2}\cdots \int_{m^2}^{k_2}\frac{d
k_1^2}{k_1^2}\int_x^1 d\b_n\Th^{(1)}_n\int_{\b_n}^1
d\b_{n-1}\Th^{(1)}_{n-1}...
\n\\
\t\int_{\beta_2}^1d \beta_1\Th^{(1)}_1
P(\frac{\beta_n}{\beta_{n-1}})...P(\beta_1), \quad
P(z)=2\frac{1+z^2}{1-z},
\end{eqnarray}
where the limits of integrals show the intervals of variation and the
integrand is the differential cross section. Again the rapidities
$\beta_i$ are arranged and the transverse momenta
square are rigorously arranged. Here
$\Th^{(i)}=\theta(\theta_{i+1}-\theta_i)$ the condition which forbids
the destructive interference of jets and jets emission angles are
$\theta_i=|\vec{k}_i|/(E\max{|\alpha_i|,|\beta_i|})$ regarding the
beam axe direction.

Compared with peripheral production regime DIS one gives the
contribution which fall down with increasing $Q^2$.  Using the
analogy to statistical nonequilibrium processes we may consider DIS
regime as an nonequilibrium process of diminishing of large
virtualities to small ones by means of evolution. The peripheral
regime may be associated with equilibrium process when the random
fluctuations on the transvers momenta values become essential.

Let consider now the regime of hard particles production at large
angles. It is known as a double-logarithm regime. Any exclusive
process is suppressed in this regime by Sudakov form factor. For
inclusive set-up of experiments when arbitrary number of photons
(gluons) may be emitted the Sudakov's form factor suppression
disappears and we obtain the cross section of the form [14]:
\begin{equation}
d\sigma(s)=\frac{1}{s}F(\alpha\ln^2\frac{s}{m^2}).
\end{equation}
For instance the cross section of annihilation of electron-positron to
muon pair accompanied by emission of arbitrary number of photons is
$$
\sigma(s)_{e\bar{e}\to \mu\bar{\mu}+...}=\frac{4\pi\alpha^2}{3s}ch x,
x^2=\frac{2\alpha}{\pi}\ln^2(\frac{s}{m_\mu m_e}).
$$
The characteristic squares of transvers momenta of created particles
in this regime are big and of order of $s$.

The typical process -annihilation of electron -positron pair to $n$ hard
photons accompanied by emission of any number of soft and virtual photons
have a form [14]:
\begin{eqnarray}
d\s_n=\frac{2\pi\alpha^2}{s}dF^{DL}_n, \\ \nonumber
dF^{DL}_n=(\frac{\alpha}{2\pi})^n\prod_{i=1}^{i=n} dx_i d
y_i\theta(y_i-y_{i-1})
\theta(x_i-y_i)\theta(x_i)\theta(y_i)\theta(\rho-x_n)\theta(\rho-x_n),
\\ \nonumber
x_i=\ln\frac{\vec{q}_i^2}{m^2},y_i=
\ln\frac{1}{\beta_i},\rho=\ln\frac{s}{m^2}.
\end{eqnarray}

\section{Various scenario}\0

Keeping in mind the kinematical restriction $\Pi s_j\sim s$ we may
build the combinations of regimes considered above.
Le construct the relevant cross sections. It is convenient
to separate them to the classes\\
a) Pomeron regime (P);\\
b) Evolution regime (DIS);\\
c) Double logarithmic regime (DL);\\
d) DIS+P regime;\\
e) P+DL+P regime.\\
The description of every regime may be performed in terms of
effective ladder-type Feynman diagrams (The set of relevant FD
depends on the gauge chosen and include much more number of them).

The Pomeron is treated as a (infinite) set of particles emitted close
to the CM beams direction (within the small angles of order
$\theta_i\sim 2m_h/\sqrt{s}<<1$). We expect that these type
of particles will not be detected by the detectors since
they are move into the beams pipe. The collider experiment detectors
locate at finite angles $\theta_D\sim 1$ and will measure the
products only of particle $c$ decay.

What will happened when instead of one particle (see (1.1)) a set of
particles with invariant mass square $s_t$ is created at large
angles? Then the cross section will have a form:
\ba
d\s_n=\f{\a_s^2}{s_t}N dF^{DL}_n(\a_s\ln^2(\f{s_t}{s_0})),~
N=(\f{s}{s_t})^\D,
\n\\
\D=\a_P-1=\f{12\ln2}{\pi}\a_s\approx0.55,~\a_s=0.2
\ea
Radiative corrections to the intercept was calculated \cite{fl98} in
recent time. The resulting value is $\Delta\approx 0.2$.

The way to obtain detected large multiplicity is to organize DIS-like
experiments, expecting the large-angle scattered hadrons in the
detectors. Large transfers momenta will be decreasing
by ordinary evolution mechanism to the value of order $m_\pi$
and then the Pomeron mechanism of peripheral scattering of the
created hadrons from the pionization region will start.

This phenomena is quite close to flea-dog model of Euhrenfest (see
[9]):  in the nonequilibrium process (DIS regime) the fluctuations are
suppressed and they take place (Pomeron regime) when the equilibrium
take place.

What the characteristic multiplicities expected from Pomeron
mechanism with the intercept exceeding unity, $\Delta\sim 0.2$? It is
the quantity of order $(s/m_\pi^2)^\Delta\approx 200$ for
$\sqrt{s}=14 TeV$. This rather rough estimation is in agreement
with the phenomenological analysis of A.Kaidalov \cite{kai}, based
on multi-pomeron exchange in the scattering channel.

Construct now the cross sections of combined processes.
When the one of the initial particles $h_1$ is scattered on small but
sufficient enough angle to fit
the detectors and other is scattered almost forward the combination
of DIS and Pomeron regimes take place:
\begin{equation}
d\sigma_{n,m}=d\sigma_n^{DIS} dZ_m,|q_n|^2\sim m^2
\end{equation}
provided that the virtuality of the last step of evolution regime of
order of hadron mass. For the kinematical case of almost forward
scattering of both initial hadrons the situation may be realized with
large angles hadron production from the central region (see (3.1)):
\begin{equation}
d\sigma_{n,m,k}=dZ_n d\sigma^{DL}_m d Z_k.
\end{equation}

\section{Discussion}

Every regime $a)-e)$ considered above provide creation of
$(n+2)$ gluons and theirs subsequent decay in the universal way on
jets.

Let us discuss more carefully the large-angle moving jets creation
mechanism due to annihilation of initial partons into the hadrons
system.  The intermediate state of single heavy photons decay into
jet was analyzed by A.Polyakov \cite{POL}.  It was shown that the
scaling regime works, providing the behavior of the $n$ jets
creation cross section and the mean multiplicity as:
\begin{equation}
\s_n(s)\sim\frac{1}{s}(\frac{m^2}{s})^\delta
F(n(\frac{m^2}{s})^\delta), \bar{n} \sim(\frac{s}{m^2})^\delta,
0<\delta<\frac{1}{2}.
\end{equation}
Another mechanism of multiple production \cite{KNO} takes into
account the channels of incident partons annihilation onto the
arbitrary large number of real and virtual gluons (photons). On this
way some intermediate regime between the single logarithm (the
renormalization group approach) and the double logarithmical,
$\alpha\ln^\rho(s/m^2)\sim 1, 1<\rho<2$, is realized.

All the considered regimes of many particles state production
describes the hard stage of process. The hadronization stage will
impose its features which can not be expressed in terms of pQCD as
well as it concern the confinement region.  Here the identity of
produced particles must be taken into account \cite{PODG}.  The
effective Lagrangian approach also may be applied here \cite{wein}.
We hope to consider last questions in another work.

\vskip 0.5cm
{\bf Acknowledgments}
\vskip 0.3cm

We are grateful to V.G.Kadyshevski for interest to discussed in the
paper questions.

\end{document}